\documentclass[12pt]{article}
\usepackage{graphicx}

\newcommand\pubnumber{}
\newcommand\pubdate{\today}

\def\Marseille{CPPM, Aix-Marseille Universit\'e, CNRS/IN2P3, Marseille, FRANCE}

\def\Title#1{\begin{center} {\Large #1 } \end{center}}
\def\Author#1{\begin{center}{ \sc #1} \end{center}}
\def\Address#1{\begin{center}{ \it #1} \end{center}}

\newcommand\pubblock{\rightline{\begin{tabular}{l} \pubnumber\\
         \pubdate  \end{tabular}}}
\newenvironment{Abstract}{\begin{quotation}  }{\end{quotation}}
\newenvironment{Presented}{\begin{quotation} \begin{center} 
             PRESENTED AT\end{center}\bigskip 
      \begin{center}\begin{large}}{\end{large}\end{center} \end{quotation}}

\def\beq{\begin{equation}}
\def\eeq#1{\label{#1}\end{equation}}
\def\eeqn{\end{equation}}

\def\beqa{\begin{eqnarray}}
\def\eeqa#1{\label{#1}\end{eqnarray}}
\def\eeqan{\end{eqnarray}}

\let\bar=\overbar

\def\Dslash{\not{\hbox{\kern-4pt $D$}}}
\def\dslash{\not{\hbox{\kern-2pt $\del$}}}

\def\ee{e^+e^-}

\def\msb{{\bar{\ssstyle M \kern -1pt S}}}

\begin{document}
\begin{titlepage}
\pubblock

\vfill
\Title{Charm semileptonic decays at the B factories}
\vfill
\Author{Justine Serrano\footnote{from the Babar Collaboration}}
\Address{\Marseille}
\vfill
\begin{Abstract}
A review of  charm semileptonic decays results obtained at B factories is presented. It focuses on form factors measurements 
 in $D^0\to K^- \ell^+ \nu$,  $D^0\to \pi^- \ell^+ \nu$, $D_s^+ \to K^+K^- e^+\nu_e$, and $D^+ \to K^-\pi^+ e^+\nu_e$. For the last two decay channels,
the contribution and characteristics of other components in the final state, in addition to the main vector contribution, is also studied.
 
\end{Abstract}
\vfill
\begin{Presented}
The 6th International Workshop on the CKM Unitarity Triangle,\\
 University of Warwick, UK, 6-10 September 2010
\end{Presented}
\vfill
\end{titlepage}
\def\thefootnote{\fnsymbol{footnote}}
\setcounter{footnote}{0}

\section{Introduction}
Detailed study of charm semileptonic decays is interesting for several reasons. The measurement of decay rates provides a
determination of the hadronic form factors entering these decays, as the CKM matrix elements $|V_{cs}|$ and $|V_{cd}|$ are precisely known. Hadronic form factors measurements in $D$ decays can help validate predictions from lattice QCD calculations, which are necessary to determine $|V_{cb}|$ and $|V_{ub}|$ from $B$ semileptonic decays. The golden modes to perform this validation are the decays 
of a $D$ into a pseudoscalar meson. The study of $D$ decays into vector states are more complicated but still interesting to get a complete understanding of charm semileptonic decays. Moreover,  $D$ semileptonic decays involving two pseudoscalar mesons in the final states allow to precisely measure  the characteristics of hadronic systems without additional hadrons, unlike three body Dalitz plot analyses.

During the past five years, B factories have demonstrated their capability in 
performing accurate measurements in the field of charm semileptonic decays.
While CLEO-c events benefit from a very clean environment, B factories have the advantage of much larger  statistics (the cross-section of $c\bar c$ events at the $\Upsilon(4S)$ energy is 1.3 nb). 
Moreover, since charm hadrons are produced by fragmentation, different flavour of charm mesons and 
baryons can be studied. 
To reconstruct signal events, BABAR and Belle have developed different techniques, which both lead to competitive results.
Belle performed a tagged analysis with  full reconstruction of the events $\ee\to D^{(*)}_{tag} D^{(*-)}_{sig}X $, where $X$ may be  additional mesons from  fragmentation.
This gives access to a  good $q^2$ resolution and to an absolute measurement of the branching fraction at the price of a low efficiency. 
In BABAR, only the signal charm meson is reconstructed, the neutrino energy being evaluated from
the rest of the event. This method has  much higher efficiency but the  branching fraction measurement 
has to be done using a normalization channel. 

\section{$D\to P \ell \nu$ decays}
Neglecting the lepton mass, the differential decay rate depends on only  one form factor $f_+(q^2)$, where $q^2$ is the invariant mass squared 
of the two leptons. The models used to parameterized the $q^2$ dependence of the form factor are the modified pole model \cite{Becirevic:1999kt}, where the parameter to be fitted is $\alpha$, and the simple pole model, for which the mass of the pole is fitted.
Using 282 fb$^{-1}$ of data, Belle studied the $D^0\to K^- \ell^+ \nu$ and $D^0\to \pi^- \ell^+ \nu$ decay channels for $\ell=\mu,e$ \cite{Widhalm:2006wz}. 
BABAR measured the form factors parameters for the  $D^0\to K^- e^+ \nu$ channel 
and the branching fraction relative to the $D^0\to K^-\pi^+$ channel using 75 fb$^{-1}$ \cite{Aubert:2007wg}.

For the $D^0\to \pi^- \ell^+ \nu$ channel, Belle measures $\alpha=0.10\pm0.21\pm 0.10$ and $f_+(0)=0.624\pm 0.020 \pm 0.030$,
which is compatible with CLEO-c and lattice QCD  determinations\cite{Aubin:2004ej}. 
The experimental precision is of the order of $6\%$ on $f_+(0)$ while 
lattice predictions are $10\%$ accurate.

As one can expect,  the $D^0\to K^- e^+ \nu$ decay properties are more precisely determined experimentally. BABAR measures $\alpha=0.38\pm0.02\pm 0.03$, with a precision similar to the CLEO-c experiment. 
Belle finds $\alpha=0.52\pm0.08\pm 0.06$, well in agreement with 
the lattice QCD value ($\alpha=0.50\pm0.04\pm0.07$).
The absolute form factors normalization are $f_+(0)=0.695\pm 0.007 \pm 0.022$ for Belle and $f_+(0)=0.735\pm 007 \pm 0.005 \pm 0.005$ for BABAR, where the third error comes from external inputs. These measurements agree with the most recent lattice QCD calculation, which has reached a 2.5$\%$ precision \cite{Na:2010uf}.
It has to be noted that all the experimental results exclude the simple pole model as they find a mass pole 
below the expected $D_s^*$ mass. 
Predictions from the popular Isgur-Wise 2 model are also excluded by these measurements.

\section{$D\to V \ell \nu$ decays}
\begin{figure}[htb]
\centering
\includegraphics[height=1.5in]{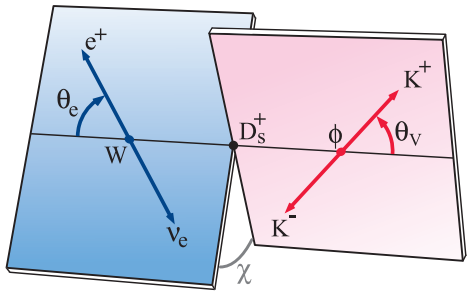}
\caption{Definition of kinematic variables for the $D_s^+ \to K^+K^- e^+\nu_e$ decay channel.}
\label{fig:dec}
\end{figure}

Semileptonic decays into a vector meson depend on five variables ($q^2$, three decay angles as shown on Figure \ref{fig:dec}, and the mass squared of the vector meson products)
 and on three form factors, $A_1$, $A_2$ and $V$, for which we assume a $q^2$ dependence
dominated by a single pole:
\begin{eqnarray*}
V(q^2)=\frac{V(0)}{1-q^2/m_V^2};~A_{1,2}(q^2)=\frac{A_{1,2}(0)}{1-q^2/m_{A}^2}. \\
\end{eqnarray*}

More generally, we  consider the $D\to PP' \ell \nu$ decays. Ususally the $PP'$ state is dominated by a vector state, but 
we also take into account other non dominant S- or P-waves.

\subsection{$D_s^+ \to K^+K^- e^+\nu_e$}

BABAR analyses $214 ~\rm{fb^{-1}}$ of data, selecting about 25000 events
with a $K^+K^-$ invariant mass around the $\phi$ region (in the range $1.01-1.03~\rm{GeV/c^2}$) \cite{Aubert:2008rs}.
This  statistics largely exceeds that of any previous measurement and allows the determination of the pole mass $m_A$ in addition to the usual form factors ratios parameters $r_2=A_2(0)/A_1(0)$, $r_V=V(0)/A_1(0)$ evaluated at $q^2=0$. These parameters
are extracted using a binned maximum likelihood fit to the four-dimension decay distribution.
The sensitivity to the vector pole mass $m_V$ is weak and this parameter is fixed to $2.1~\rm{GeV/c^2}$. The following 
values are obtained: $r_2=0.763\pm0.071\pm0.065$, $r_V=1.849\pm0.060\pm0.095$, $m_A=2.28^{+0.23}_{-0.18}\pm0.18 ~\rm{GeV/c^2}$.  
Measuring the $D_s^+ \rightarrow K^+ K^- e^+ \nu_e$ branching fraction
relative to the decay $D_s^+ \rightarrow  K^+ K^- \pi^+$, the absolute normalization
is obtained and $A_1(0) = 0.607 \pm 0.011 \pm 0.019 \pm 0.018$.
Lattice  calculations for this channel have
been done in the quenched approximation\cite{Gill:2001jp}. They agree with the present experimental result
of  $A_1(0)$, $r_2$ and $m_A$, but are lower than the measured value 
of $r_V$.

BABAR also finds a first evidence for a small S-wave contribution in the analysed mass range, through its interference with the $\phi$, which may be associated 
with $f_0 \to K^+K^-$ decays and corresponds to $(0.22^{+0.12}_{-0.08}\pm0.03)\%$ of the
$K^+K^- e^+\nu$ decay rate. With this hypothesis and extrapolating to the total mass range using the $f_0$ parameters measured by BES \cite{Ablikim:2004wn}, the ratio of S- and P-wave contributions is:
\begin{eqnarray*}
\frac{BR(D_s^+ \to f_0^- e^+\nu_e)}{BR(D_s^+ \to \phi e^+\nu_e)}= (4.5^{+2.5}_{-1.6}\pm0.6)\%.
\end{eqnarray*}

\subsection{$D^+ \to K^-\pi^+ e^+\nu_e$}
This decay has been studied by BABAR using $347.5 ~\rm{fb^{-1}}$ of data \cite{Sanchez:2010fd}. They select 244 $\times 10^{3}$ signal events which allow
to perform a fit to the five-dimension decay distribution in the whole phase space, including the contributions from 
the S-wave, the $K^*(892)$, and the $K^*(1410)$. 
The form factors parameters associated to the $K^*(892)$ are accurately measured\footnote{$m_V$ is fixed as in the $D_s^+ \to K^+ K^- e^+\nu_e$ analysis}: 
$r_2=0.801\pm0.020\pm0.020$, $r_V=1.463\pm0.017\pm0.031$, $m_A=2.63\pm0.10\pm0.13~\rm{GeV/c^2}$.  
Measuring the $D^+ \to K^-\pi^+ e^+\nu_e$ branching fraction
relative to the decay $D^+ \rightarrow  K^- \pi^+ \pi^-$, the absolute normalization
is obtained and $A_1(0) = 0.6226 \pm 0.0056 \pm 0.0065 \pm 0.0074$.
The mass, width, and Blatt-Weisskopf parameters of the $K^*(892)$ are also measured.

In addition to the $K^*(892)$ meson, BABAR finds a contribution of the $ K^- \pi^+$ S-wave component of $(5.79\pm 0.16\pm0.15)\%$, 
which is comparable to the one found in the $D_s^+ \rightarrow K^+ K^- e^+ \nu_e$ channel, and a small contribution from 
the $K^*(1410)$ equal to ($0.33\pm 0.13\pm0.19)\%$.

The $K^- \pi^+$ S-wave phase is measured in bins of the $ K^- \pi^+$ mass and is in agreement with measurements obtained in $K^- \pi^+$
production at small momentum transfer in fixed target experiments, with an opposite sign between the S- and P-wave.

\section{Conclusion}
Thanks to the large statistics of charm mesons produced at the $\Upsilon(4S)$ and to original analysis techniques, 
B factories have been able to reach  high accuracy in $D$ semileptonic decays measurements. Performances similar to dedicated
charm experiments have been obtained in $D\to P \ell \nu$ decays and unprecedent signal yield obtained for 
$D\to PP' \ell \nu$ decays have lead to precise form factors determinations and detailed hadronic system studies.


\end{document}